\documentclass[twocolumn,prl]{revtex4}

\usepackage{graphicx}
\usepackage{dcolumn}
\usepackage{bm}
\usepackage{color}

\begin{document}

\preprint{}

\title{Quantum Phase Transitions in the Hubbard Model on Triangular Lattice}
\author{Takuya Yoshioka$^{1}$}
\author{Akihisa Koga$^{2}$}
\author{Norio Kawakami$^{2}$}
\affiliation{$^{1}$Department of Applied Physics, Osaka University, Suita, Osaka, 565-0871, Japan\\
$^{2}$Department of Physics, Kyoto University, Kyoto, 606-8502, Japan}

\date{\today}

\begin{abstract}
We investigate the quantum phase transitions in the half-filled Hubbard model on the triangular lattice by means of the path-integral renormalization group (PIRG) method with a new iteration scheme proposed recently. It is found that as the Hubbard interaction $U$ increases, the paramagnetic metallic state undergoes a first-order phase transition to a nonmagnetic insulating (NMI) state at $U_{c1}\sim 7.7t$, which is followed by another first-order transition to a $120^\circ$ N\'eel ordered state at $U_{c2}\sim 10t$, where $t$ is the transfer integral. Our results ensure the existence of the intermediate NMI phase, and resolve some controversial arguments on the nature of the previously proposed quantum phase transitions. We find that $\kappa$-(BEDT-TTF)$_2$Cu$_2$(CN)$_3$ is located in the NMI phase close to the metal-insulator transition point.

\end{abstract}

\maketitle

Strongly correlated electron systems with frustration have attracted much interest recently. There are a number of intriguing phenomena that have revealed new aspects of electron correlations. One of the striking examples can be found in an organic compound $\rm \kappa-(BEDT-TTF)_2Cu_2(CN)_3$ \cite{exp0,exp1,exp2,exp3}, for which the triangular lattice structure of dimerized BEDT-TTF molecules plays an invaluable role in stabilizing a nonmagnetic spin-liquid insulating state down to $20$mK \cite{exp3}. The nonmagnetic insulating (NMI) state, which is totally different from the naively expected $120^\circ$ N\'eel ordered state (120N\'eel) \cite{Heisenberg1,Heisenberg2}, poses an interesting and challenging problem in the Mott transition of strongly correlated electrons on the frustrated triangular lattice \cite{Fukuyama2,Imai,Morita,Parcollet,Lee,Kyung,Aryanpour,Koretsune,Senthil,Sahebsara,Ohashi,Watanabe,Inaba}. In particular, it has been a central issue to figure out whether such a NMI state is really stabilized on the triangular lattice, and if so, how we can theoretically describe the Mott transition without any kind of magnetic ordering. The answer to the question should certainly provide us with deeper understanding of the Mott transition with strong frustration. 

Low-energy properties of such frustrated organic compounds may be described by the single-band Hubbard model on the triangular lattice at half filling \cite{exp0,Fukuyama2}. 
There have been a number of theoretical investigations on quantum phase transitions of the model \cite{Fukuyama2,Imai,Morita,Parcollet,Lee,Kyung,Aryanpour,Koretsune,Senthil,Sahebsara,Ohashi,Watanabe,Inaba}. However, most of conventional mean-field and variational treatments  fail to describe the NMI phase, suggesting that the system  may prefer the 120N\'eel state. Among those intensive studies, the pioneering work by means of the path integral renormalization group (PIRG) method \cite{Morita} is believed to provide the most reliable results since it can fully incorporate quantum fluctuations on the basis of an unbiased scheme. By this method, Morita {\it et al.} reached the remarkable conclusion that the Mott transition occurs from the metallic state to the NMI state \cite{Morita}, and the corresponding transition may be continuous. Although the conclusion shed new light on the Mott transition, there still remain some serious questions/points: (i) Is such a  continuous transition really possible in the fully frustrated system? (ii) The obtained critical value of the Hubbard interaction seems much smaller than the values deduced by other methods (see below). (iii) It is difficult to study the magnetic instability to the 120N\'eel phase by the naive PIRG method, so that it is unclear whether the NMI phase is indeed realized against the magnetic instability. We note that a more recent variational-cluster study found the metallic, NMI and 120N\'eel phases in a certain parameter regime, but the transition points could not be estimated due to the problem of its numerical accuracy \cite{Sahebsara}. Therefore, it is highly desirable to precisely determine the ground-state phase diagram and clarify the nature of the associated quantum phase transitions in order to elucidate the essential properties inherent in the Mott transition under strong frustration. 


In this paper, we investigate the quantum phase transitions in the half-filled Hubbard model on the triangular lattice. By means of the PIRG method \cite{PIRG1,PIRG2} with an improved iteration scheme proposed recently \cite{Yoshioka}, we discuss how the NMI state competes with the metallic and 120N\'eel states. By computing the double occupancy, the momentum distribution function and the spin/charge correlation functions, we find that there are two successive first-order quantum phase transitions among the metal-NMI-120N\'eel phases, for which the transition points are determined precisely. The present results substantially improve the previous PIRG phase diagram \cite{Morita}, and shed light on some controversial arguments on the nature of the quantum phase transitions. 

\begin{figure}
  \begin{center}
      \resizebox{80mm}{!}{\includegraphics{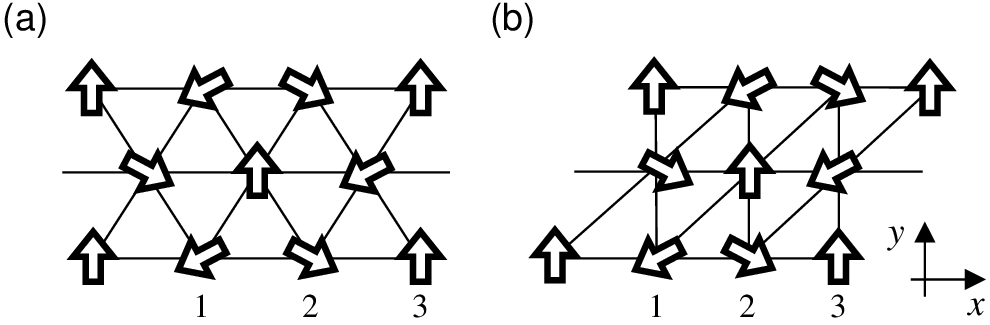}} 
  \end{center}
\caption{(a) Triangular lattice and 
(b) square lattice with diagonal transfers, 
which is topologically equivalent to the original lattice (a). 
The arrows represent the spin configuration for the 120N\'eel state  
and the associated indices specify three different  sublattices.
}
\label{fig1}
\end{figure}

Let us consider the single-band Hubbard model on the triangular lattice. For simplicity, we adopt the square lattice with some diagonal bonds as shown in Fig.~\ref{fig1}, which is topologically equivalent to the triangular lattice.  The Hamiltonian we consider reads
\begin{equation}
\hat{\cal H}=-t\sum_{\langle i,j\rangle, \sigma}
\hat{c}^{\dagger}_{i\sigma}\hat{c}_{j\sigma}
+U\sum_{i}\hat{n}_{i\uparrow}\hat{n}_{i\downarrow},
\label{H}
\end{equation}
where $\hat{c}_{i\sigma}$ ($\hat{c}^{\dagger}_{i\sigma}$) is an annihilation (creation) operator of an electron at the $i$th site with spin $\sigma(=\uparrow, \downarrow)$ and $\hat{n}_{i\sigma}= \hat{c}^{\dagger}_{i\sigma}\hat{c}_{i\sigma}$. $U(>0)$ is the Hubbard repulsion and $t(>0)$ is the nearest-neighbor transfer integral.

To study the ground-state properties of the Hubbard model eq.~(\ref{H}), we use the PIRG method developed by Imada group \cite{PIRG1,PIRG2}. The method is based on a simple idea that the true ground-state  $|\psi_g\rangle$ is obtained by acting the imaginary-time evolution operator on an initial state $|\phi_0\rangle$: $|\psi_g\rangle = e^{-\beta \hat{\cal H}}|\phi_0\rangle$ 
with $\beta\rightarrow\infty$.
To approach the true ground-state in the PIRG method, we operate $e^{-\Delta\tau \hat{\cal H}}$ with small $\Delta\tau$ on an approximate ground-state iteratively. 
This method is, in principle, independent of an initial state and an iterative scheme employed. However, how to choose them are crucial to reach the correct ground-state within restricted numerical resources. Here we use a new iteration scheme proposed in our previous paper \cite{Yoshioka}, which is extremely efficient in performing the PIRG calculations with a large number of basis states.  We examine various initial states deduced from  unrestricted Hartree-Fock (UHF) solutions \cite{UHF}. Besides, we further consider another class of initial states derived from UHF solutions in the spin-rotated frame where the quantization axis of spin is rotated with angles $\theta_0, \theta_0+2\pi/3,$ and $\theta_0+4\pi/3$ respectively for three different sublattices (see Fig.~\ref{fig1}). 
\begin{figure}[htb]
  \begin{center}
\includegraphics[width=0.38\textwidth]{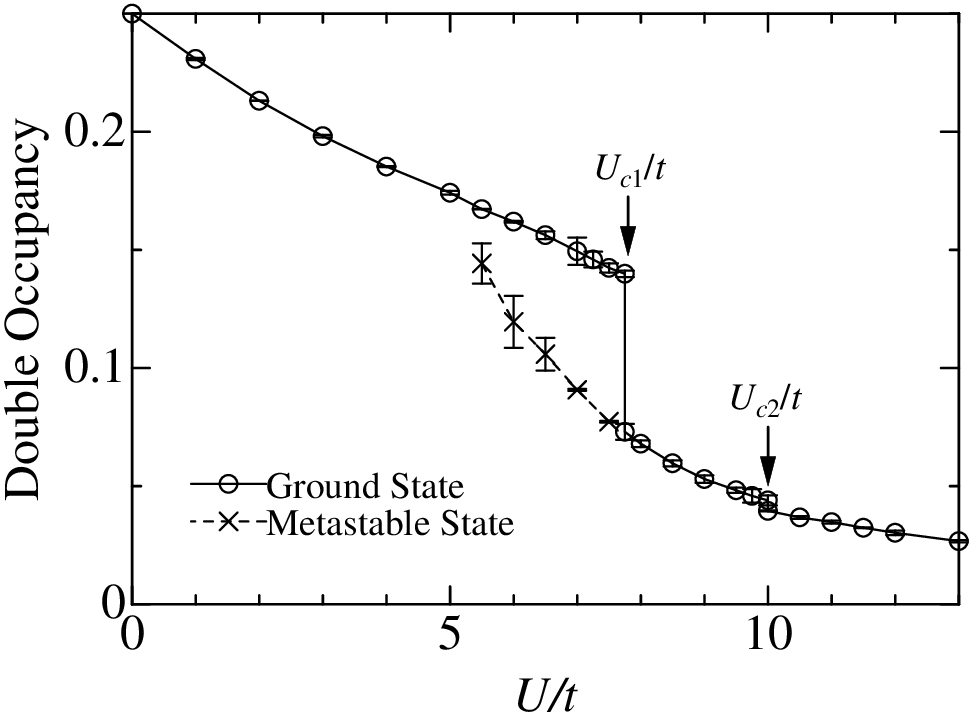}
\caption{
The double occupancy as a function of $U/t$ for the half-filled Hubbard model on the triangular lattice with $N=6\times6$ sites. Circles (crosses) represent the results for the ground (metastable) state.
}
\label{fig2}
\end{center} 
\end{figure}
Performing the PIRG calculation with the initial states mentioned above, we discuss the ground-state properties of the Hubbard model on the triangular lattice. In this paper, we focus on the system of $6\times 6$ sites with periodic boundary conditions. We have checked in a slightly different but related 2-dimensional frustrated lattice model that the system with $6\times 6$ sites provides reliable results on the quantum phase transitions \cite{Yoshioka}.
In our PIRG calculations, we keep a large number of states (up to 500) as the Slater basis states, and fix $\Delta\tau/U=0.5$.

\begin{figure}[htb]
  \begin{center}
\includegraphics[width=0.38\textwidth]{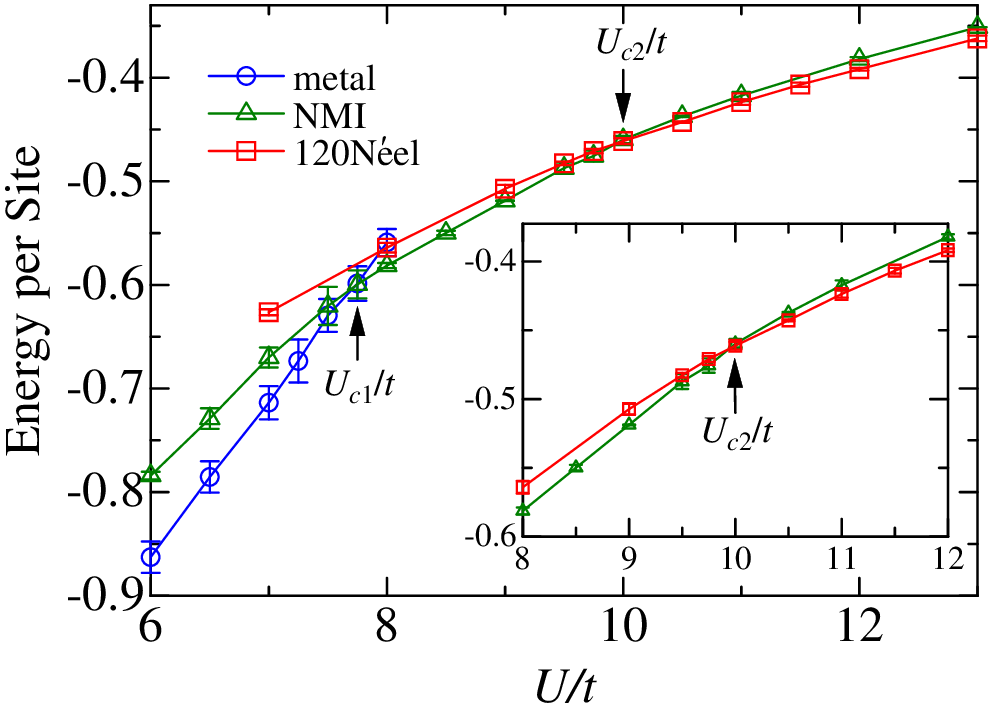}
\caption{
(color online)
The energies per site for the competing states in the half-filled Hubbard model on the triangular lattice, where circles, triangles, and squares respectively represent the energies for the nonmagnetic metal, the NMI and  the 120N\'eel.
}
\label{fig3}
\end{center} 
\end{figure}
    \label{fig4}
\begin{figure*}[htb]
\begin{tabular}{cp{0mm}cccp{3mm}p{60mm}}
      &&$U/t=6$&$U/t=9$&$U/t=12$&&\\
\vspace{3mm}\\
      \raisebox{13mm}{$n({\mbox{\boldmath$q$}})$}
      &&
      \includegraphics[width=30mm]{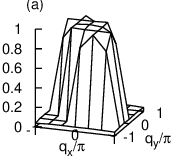} & 
      \includegraphics[width=30mm]{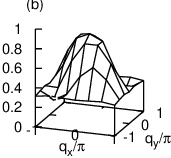} &
      \includegraphics[width=30mm]{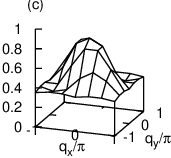} &&\\
      \raisebox{13mm}{$N({\mbox{\boldmath$q$}})$}
      &&
      \includegraphics[width=30mm]{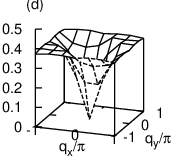} &
      \includegraphics[width=30mm]{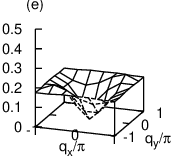} &
      \includegraphics[width=30mm]{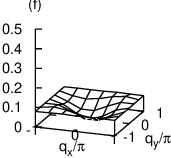} &&\\
      \raisebox{13mm}{$S({\mbox{\boldmath$q$}})$}
      &&
      \includegraphics[width=30mm]{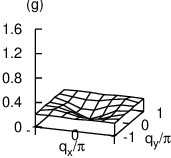} & 
      \includegraphics[width=30mm]{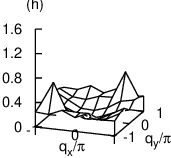} &
      \includegraphics[width=30mm]{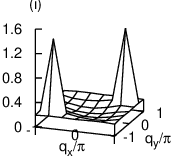} &&
\vspace{-70mm}\caption{
The momentum distribution function $n({\mbox{\boldmath$q$}})$, and 
the momentum-dependent charge [spin] correlation function 
$N({\mbox{\boldmath$q$}})$ [$S({\mbox{\boldmath$q$}})$] 
for the half-filled Hubbard model on the square lattice with $U/t=6, 9$ and $12$. For these values of interaction, the system is respectively in the metallic, NMI and 120N\'eel phases. Note that the charge correlation shows the metallic behavior $N({\mbox{\boldmath$q$}}) \sim |{\mbox{\boldmath$q$}}|$ in (d) and the insulating behavior $N({\mbox{\boldmath$q$}}) \sim |{\mbox{\boldmath$q$}}|^2 $ in (e) and (f).
    \label{fig4}
}
    \end{tabular}
\end{figure*}
We first compute the expectation value of the double occupancy $\sum_{i=1}^N\langle\hat{n}_{i\uparrow}\hat{n}_{i\downarrow}\rangle/N$. The results are shown in Fig.~\ref{fig2}. The introduction of the repulsive interaction monotonically decreases the double occupancy, implying that the highly correlated metallic state is realized for $U<U_{c1}$. Further increase in the interaction gives rise to two successive discontinuities in the curve at $U=U_{c1}$ and $U_{c2}$ although the latter singularity is rather weak. We thus find that double first-order quantum phase transitions occur in the system. This is also supported by the appearance of the two cusp singularities in the curve of the ground-state energy, as shown in Fig.~\ref{fig3}.
By estimating the level crossing points of energies for the competing states, we determine the transition points $U_{c1}\sim 7.7t$ and $U_{c2}\sim 10t$.

Having uncovered that both of the quantum phase transitions are of first order, let us now discuss the nature of the three distinct phases in detail. To this end, we calculate the momentum distribution function $n({\mbox{\boldmath$q$}})$ and the momentum-dependent correlation function in the charge [spin] sector $N({\mbox{\boldmath$q$}})$ $[S({\mbox{\boldmath$q$}})]$.  These quantities are respectively given by the Fourier transform of the site-dependent correlation functions, $\langle \hat{c}^{\dagger}_{i\sigma}\hat{c}_{j\sigma}\rangle$, 
$\left(\langle \hat{n}_{i}\hat{n}_{j}\rangle-\langle \hat{n}_{i}\rangle\langle \hat{n}_{j}\rangle\right)$ 
and 
$\langle \hat{\mbox{\boldmath$S$}}_{i}\cdot 
\hat{\mbox{\boldmath$S$}}_{j}\rangle$,
where $\hat{n}_i=\sum_\sigma \hat{n}_{i\sigma}$, 
$\hat{\mbox{\boldmath$S$}}_{i}=1/2 \sum_{\alpha\beta}
\hat{c}^{\dagger}_{i\alpha} \sigma_{\alpha\beta} \hat{c}_{i\beta}$ and
$\sigma$ is the Pauli matrix. The computed results are shown in Fig.~\ref{fig4}. For $U < U_{c1}$, we find that the discontinuity exists in the momentum distribution function at the Fermi surface (Fig.~\ref{fig4} (a)). In this case, no singularity appears in the charge and spin correlation functions in Figs.~\ref{fig4} (d) and (g). We thus confirm that the ordinary paramagnetic metallic state is stabilized for $U<U_{c1}$. On the other hand, when $U>U_{c1}$, the jump singularity disappears in the momentum distribution function, as shown in Figs.~\ref{fig4} (b) and (c), in accordance with the Mott transition at $U=U_{c1}$. Correspondingly, the charge correlation function $N({\mbox{\boldmath$q$}})$ changes its $|{\mbox{\boldmath$q$}}|$-dependence (small $|{\mbox{\boldmath$q$}}|$ region) from linear to quadratic. In the region $U_{c1}<U<U_{c2}$, the repulsive interaction enhances spin fluctuations at ${\mbox{\boldmath$q$}}_{peak}=(\pm 2\pi/3, \pm 2\pi/3)$ characteristic of the 120N\'eel phase, but does not give rise to divergent behavior. 

\begin{figure}[htb]
  \begin{center}
\includegraphics[width=0.38\textwidth]{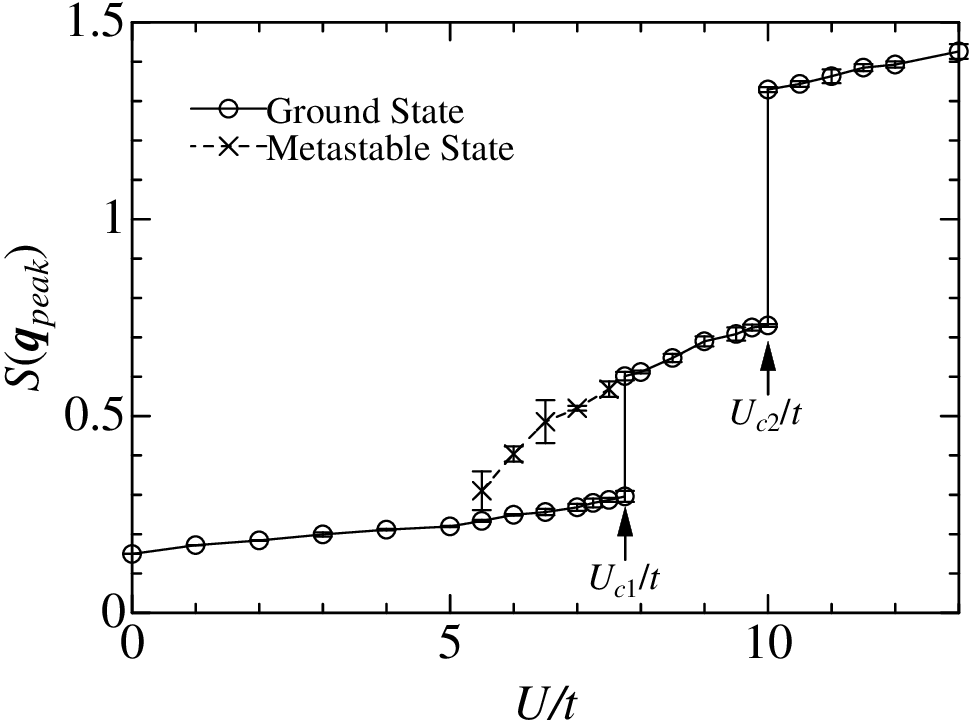} 
\caption{
The spin correlation function 
$S({\mbox{\boldmath$q$}}_{peak})$ 
as a function of $U/t$ on the $N=6\times6$ lattice 
at half filling for ${\mbox{\boldmath$q$}}_{peak}$.
Circles (crosses) represent the results for the ground (metastable) state.
}
\label{fig5}
  \end{center}
\end{figure}

To discuss how magnetic fluctuations induce the 120N\'eel phase for $U>U_{c2}$, we plot $S({\mbox{\boldmath$q$}}_{peak})$ as a function of $U/t$ in Fig.~\ref{fig5}. We find two clear jumps at $U_{c1}$ and $U_{c2}$ that signal the first-order phase transitions. It is to be noted that spin fluctuations are strongly enhanced for $U>U_{c2}$, suggesting the emergence of the 120N\'eel phase. Although we have not performed the finite-size scaling in terms of different system sizes here, the tendency is consistent with the well-known fact for the triangular-lattice Heisenberg model \cite{Heisenberg1,Heisenberg2}. These observations naturally lead us to the conclusion that 120N\'eel phase is realized for $U>U_{c2}$. On the other hand, in the intermediate phase $U_{c1}<U<U_{c2}$, the long-range order for the $120^\circ$ spin configuration is destroyed, where $S({\mbox{\boldmath$q$}}_{peak})$ does not show significant size-dependence. Therefore, we can say that the NMI state is indeed realized between the metallic and 120N\'eel phases. 

It should be noted that the present results are in stark contrast to the previous PIRG results of Morita {\it et al}. \cite{Morita}. First, our numerical data clearly demonstrate the first-order transition between the metal and insulator, while  Morita {\it et al.} claimed that the transition may be continuous. Furthermore, the transition point $U_{c1}\sim 7.7t$ largely exceeds $U_c\sim 5.2 t$ obtained in the latter work \cite{Morita}. Secondly, our new PIRG algorithm can describe the 120N\'eel insulating phase which could not be treated in the previous study. Here we argue what really causes the difference between these PIRG results. In general, it may be difficult to discuss first-order transitions by iterative schemes since several competing states should be treated precisely on an equal footing. We find that the PIRG calculation with a naive iteration scheme may fail to reach the correct ground-state, since the iteration procedure suffers from metastable states around the metal-insulator transition point. We show a typical example in Figs.~\ref{fig2} and  \ref{fig5}. Suppose we regard such a metastable state as the ground-state, the cusp singularity without discontinuity might appear in the curve of the double occupancy, as shown in Fig.~\ref{fig2}. This would cause a misleading conclusion that the continuous Mott transition occurs around a small critical value $U_c\sim 5 t$, which is analogous to the statement of Ref.~\cite{Morita}. Our new iteration scheme can cope with such metastable states properly \cite{Yoshioka}, giving a clear description of the first-order phase transition at $U=U_{c1}$ in Figs.~\ref{fig2}, \ref{fig3} and \ref{fig5}. This is also the case for another phase transition to the 120N\'eel phase, for which further elaborated calculations have been performed in terms of the spin-rotated basis states mentioned above. Consequently we find the first-order transition at $U=U_{c2} \sim 10t$.

\begin{figure}[htb]
  \begin{center}
\includegraphics[width=0.38\textwidth]{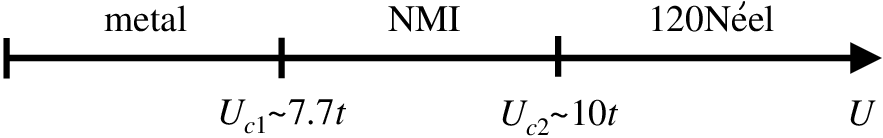}  
\caption{
Phase diagram for the half-filled Hubbard model on the triangular lattice. Both of the transitions are of first order.
}
\label{fig6}
  \end{center}
\end{figure}
We show our PIRG phase diagram in Fig.~\ref{fig6}, which substantially improves the previous one \cite{Morita} and thus sheds light on the controversial arguments on the nature of the quantum phase transitions: The metal-insulator transition is not continuous but of first order, and the corresponding transition point $U_{c1} \sim 7.7t$  is much larger than the previous one $ U_{c1}\sim 5.2$; The NMI phase proposed in Ref.~\cite{Morita} is indeed realized against the magnetic instability to the 120N\'eel phase.  

Before concluding the paper, we briefly comment on the nonmagnetic insulating state found in the organic compound $\kappa$-(BEDT-TTF)$_2$Cu$_2$(CN)$_3$. According to the band structure calculation \cite{exp0}, the transfer integral and the Coulomb interaction in the compound are estimated as $t\sim54.5$ meV, $t'\sim57.5$ meV, and $U\sim448$ meV ($U/t\sim8.2$ and $t'/t\sim1.06)$, so that the system is well described by the isotropic triangular lattice model.  By exploiting these values, we conclude that the above compound with $U/t\sim8.2$ is indeed in the NMI phase. Interestingly, the phase diagram in Fig.~\ref{fig6} tells us that the compound is located in the NMI phase close to the first-order Mott transition point. 

In summary, we have obtained the ground-state phase diagram of the Hubbard model on the triangular lattice by means of the PIRG method with the improved iteration scheme. Our analysis has uncovered that there indeed exists the NMI phase between the metallic and 120N\'eel phases. The phase diagram obtained has resolved some apparently controversial conclusions on this issue. 

\begin{acknowledgments}
This work was partly supported by the Grant-in-Aid for Scientific Research 
[19014013, 20029013 (N.K.) and 20740194 (A.K.)] and 
the Global COE Program "The Next Generation of Physics, Spun 
from Universality and Emergence" from 
the Ministry of Education, Culture, Sports, Science and Technology (MEXT) 
of Japan. 
\end{acknowledgments}


\begin{thebibliography}{00}

\bibitem{exp0}
T. Komatsu, N. Matsukawa, T. Inoue, and G. Saito, J. Phys. Soc. Jpn. {\bf 65}, 1340 (1996).

\bibitem{exp1}
Y. Shimizu, K. Miyagawa, K. Kanoda, M. Maesato, and G. Saito, Phys. Rev. Lett. {\bf 91}, 107001 (2003).

\bibitem{exp2}
Y. Kurosaki, Y. Shimizu, K. Miyagawa, K. Kanoda, and G. Saito, Phys. Rev. Lett. {\bf 95}, 177001 (2005).

\bibitem{exp3}
S. Ohira, Y. Shimizu, K. Kanoda, and G. Saito, J. Low Temp. Phys. {\bf 142},153 (2006).

\bibitem{Heisenberg1}
B. Bernu, C. Lhuillier, and L. Pierre, Phys. Rev. Lett. {\bf 69}, 2590 (1992).

\bibitem{Heisenberg2}
L. Capriotti, A. E. Trumper, and S. Sorella Phys. Rev. Lett. {\bf 82}, 3899 (1999).


\bibitem{Fukuyama2}
H. Kino and H. Fukuyama, J. Phys. Soc. Jpn. {\bf 65}, 2158 (1996).

\bibitem{Imai}
Y. Imai and N. Kawakami, Phys. Rev. B {\bf 65}, 233103 (2002).

\bibitem{Morita}
H. Morita, S. Watanabe, and M. Imada, J. Phys. Soc. Jpn. {\bf 71}, 2109 (2002).

\bibitem{Parcollet}
O. Parcollet and G. Biroli, and G. Kotliar Phys. Rev. Lett. {\bf 92}, 226402 (2004).

\bibitem{Lee}
S. S. Lee and P. A. Lee, Phys. Rev. Lett. {\bf 95} 036403 (2005).

\bibitem{Kyung}
B. Kyung and A. M. S. Tremblay, Phys. Rev. Lett. {\bf 97}, 046402 (2006).

\bibitem{Aryanpour}
K. Aryanpour, W. E. Pickett, and R. T. Scalettar, Phys. Rev. B {\bf 74}, 085117 (2006)

\bibitem{Koretsune}
T. Koretsune, Y. Motome, and A. Furusaki, J. Phys. Soc. Jpn. {\bf 76}, 074719 (2007).

\bibitem{Senthil}
T. Senthil, Phys. Rev. B {\bf 78}, 045109 (2008).

\bibitem{Ohashi}
T. Ohashi, T. Momoi, H. Tsunetsugu, and N. Kawakami, 
Phys. Rev. Lett. {\bf 100}, 076402 (2008). 

\bibitem{Watanabe}
T. Watanabe, H. Yokoyama, Y. Tanaka, and J. Inoue, Phys. Rev. B {\bf 77}, 214505 (2008).

\bibitem{Inaba}
K. Inaba, A. Koga, S. Suga, and N. Kawakami, cond-mat/0809.2383. (J. Phys.: Condens. Matter in press.)

\bibitem{Sahebsara}
P. Sahebsara and D. S\'en\'echal, Phys. Rev. Lett. {\bf 100}, 136402 (2008).

\bibitem{PIRG1} 
M. Imada and T. Kashima, J. Phys. Soc. Jpn. {\bf 69}, 2723 (2000).

\bibitem{PIRG2}
T. Kashima and M. Imada, J. Phys. Soc. Jpn. {\bf 70}, 2287 (2001).

\bibitem{Yoshioka}
T. Yoshioka, A. Koga, and N. Kawakami, J. Phys. Soc. Jpn. {\bf 77}, 104702 (2008).


\bibitem{UHF}
N. Furukawa and M. Imada, J. Phys. Soc. Jpn. {\bf 60}, 3669 (1991).









\end{thebibliography}
\end{document}